\documentclass{article}
\usepackage{spconf}
\usepackage{amsmath}
\usepackage{graphicx}
\usepackage{array}
\usepackage{url}
\usepackage{multirow}
\usepackage{balance}
\usepackage{bbm}
\usepackage{cite}
\usepackage{xcolor}
\usepackage{caption}
\usepackage{subfigure}
\usepackage{setspace}
\usepackage[T1]{fontenc}
\usepackage{amsfonts}

\title{Prompt-driven Target Speech Diarization}

\name{Yidi Jiang$^{1}$, Zhengyang Chen$^{2}$, Ruijie Tao$^{1*}$\thanks{* Corresponding Author}, Liqun Deng$^{3}$, Yanmin Qian$^{2}$ and Haizhou Li$^{5,4,1}$}

\address{
  $^{1}$National University of Singapore, Singapore~~~~~~$^{2}$Shanghai Jiao Tong University, China\\
  $^{3}$Huawei Noah's Ark Lab, China~~~~~~$^{4}$Shenzhen Research Institute of Big Data, Shenzhen, China \\
  $^{5}$School of Data Science, The Chinese University of Hong Kong, Shenzhen, China}
 
\begin{document}
\ninept
\topmargin=0mm
\maketitle

\begin{abstract} 
We introduce a novel task named `target speech diarization', which seeks to determine `when target event occurred' within an audio signal. We devise a neural architecture called Prompt-driven Target Speech Diarization (PTSD), that works with diverse prompts that specify the target speech events of interest. 
We train and evaluate PTSD using sim2spk, sim3spk and sim4spk datasets, which are derived from the Librispeech. We show that the proposed framework accurately localizes target speech events.
Furthermore, our framework exhibits versatility through its impressive performance in three diarization-related tasks: target speaker voice activity detection, overlapped speech detection and gender diarization. In particular, PTSD achieves comparable performance to specialized models across these tasks on both real and simulated data. This work serves as a reference benchmark and provides valuable insights into prompt-driven target speech processing.
\end{abstract}

\section{Introduction}
\label{sec:1}
%background

%\textcolor{red}{(to Yidi: this paper is poorly written. Major revision is required. 1/ please change semantic concept to semantic attribute, and `category' to `value' - meaning that each attribute takes multiple values. 2/ it is not clear what is the learning label of the data, in multiple places, a query vector is mentioned, but no where query vector is defined? doesn't it mean a frame can have multiple labels? and the multiple labels form a vector? 3/ the experiments are unclear, during training, what are the training inputs? what are the learning target? in Eq(2), we have two symbols y and d, however, in other places, we say that the learning target is the vector $p$? how does the labels of the training data look like? are they labeled at frame level?)}
Humans have the ability to selectively attend to a specific sound source in a complex acoustic environment, that is commonly referred to as the cocktail party effect~\cite{cherry1953some}. Such remarkable auditory attention mechanism allows us to focus our listening effectively~\cite{fritz2007auditory, mesgarani2012selective}. 
Speaker extraction does this when the attended target speaker is known in advance~\cite{li23ja_interspeech,xu2020spex,ge2020spex+}. Speaker diarization seeks to demarcate `who spoke when' in a multi-talker speech~\cite{chen2023attention,cheng2023target,medennikov2020target}. They serve as the front-end for several speech downstream tasks~\cite{gao2022automatic,liu2023enhancing,jiang2021knowledge}.
However, beyond speaker identity~\cite{desplanques2020ecapa,jiang2023target,liu2023disentangling,liu2023golden}, we are also interested in other semantic types of human speech~\cite{wang2023speech,tzinis2022heterogeneous,lebourdais2022overlapped}, for example, female speech, multi-talker speech mixture, or the speech of keynote speaker who speaks the most in a meeting.

\begin{figure}[!t]
    \centering
      \includegraphics[scale=0.85]{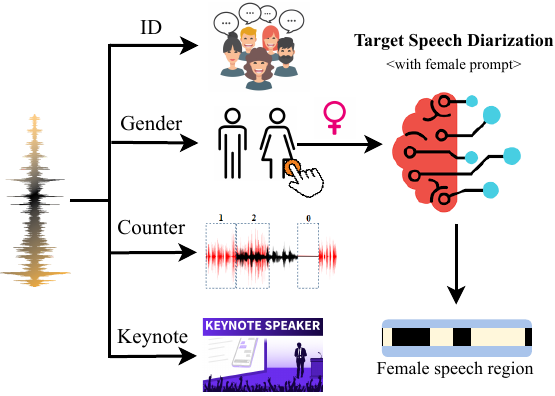}
    \caption{The illustration of the target speech diarization task. Each semantic attribute~(e.g., gender) takes on one or multiple semantic values~(e.g., female). The task aims to identify the target event regions given the semantic value information.}
    \label{fig:illustration}
\end{figure}

%define task; goal;
In this paper, we introduce a novel task, termed `target speech diarization', which seeks to determine `when target event occurred' guided by a specific prompt within an audio. From application point of view, this is similar to speech information retrieval where we use a prompt query to retrieve relevant speech segments. From speech processing point of view, this is similar to a speaker extraction task except that the prompt can be specified by a set of speech properties, referred to as semantic attributes, beyond speaker identity. We present a prompt-driven target speech diarization framework that utilizes prompt vectors to provide contextual information as the query. The proposed model architecture is inspired by similar ideas in image segmentation~\cite{kirillov2023segment,Xinyin23DeepCache,Gongfan23DepGraph}, audio source separation~\cite{liu2023separate,hu2022hierarchic}, and speech separation~\cite{tzinis2022heterogeneous}.

%our proposal
%\textcolor{red}{ (to Yidi: when we define a vector, we need to tell the dimension, and what values each dimension can take. )}
Our framework considers four semantic attributes: time-stamped speaker identity, gender, speaker counter~(number of speakers at each frame), and keynote speaker~(the most talkative speaker). Each attribute can take on one or multiple values, specifying distinct target speech events. We associate each semantic value with its respective prompt vector.
By manipulating the combination of semantic attributes, as reflected in the prompts, the proposed model allows us to search over the speech content, thus facilitating a wide range of speech applications.

The contribution of this paper can be summarised as follows:
\begin{enumerate}
    \item We introduce the innovative task of target speech diarization. 
    Here, we utilize diverse semantic attributes to distinguish different speech events, aligning with human perception and cognitive speech processes. 
    \item We propose an efficient prompt-driven target speech diarization architecture, effectively detecting target event regions by incorporating heterogeneous prompts query vectors. Meanwhile, we conduct comprehensive experiments to demonstrate the system's robustness.
    \item Our framework extends versatility to target speaker activity detection, overlapped speech detection and gender diarization, each customized to distinct semantic attribute. Notably, we also conduct comparative analysis with specialized models on both real and simulated datasets across these tasks.
\end{enumerate}

\section{Target Speech Diarization}
\label{sec:2}
\subsection{Task formulation}
\label{ssec:task_formulation}

To formulate our task, we first introduce two concepts, the semantic attribute and semantic value. The semantic attribute represents the criterion of demarcating speech segments. Each semantic attribute takes on one or multiple semantic values associated with specific events. For examples, in speaker diarization task, speaker identity is the semantic attribute. The specific speaker ID is semantic value and his/her speaking activity is the aligned speech event. 

As we mentioned in Section \ref{sec:1}, an audio can be characterized by various semantic attributes. In our proposed target speech diarization task, the system will simultaneously take audio and semantic value information as inputs and output target event regions related to this semantic value. For example, if the semantic value is female from the semantic attribute gender, the system should output the entire female-speaking regions. To demonstrate the feasibility of the task, we consider four semantic attributes in our work, denoted as $\mathcal{T}$, $\mathcal{G}$, $\mathcal{N}$ and $\mathcal{K}$:

\begin{itemize}
    \item 
    \textbf{$\mathcal{T}$ - Timestamped speaker identity:} 

In this attribute, the attribute values consist of timestamps that point to individual speakers. We use the brief timestamp-based description ``the person who spoke at the particular time'' to specify the speaker identity. 

Compared with traditional approaches which always rely on the pre-enrolled speaker embeddings, our system is flexible and user-friendly for real-world applications. For instance, when we seek to locate all speech segments for a person of interest, we can simply scan the audio to find a timestamp when he/she is the only one talking.

    \item \textbf{$\mathcal{G}$ - Gender:} 
    
    The gender attribute is more straightforward and contains two values, female and male, which can guide the system to output the gender-specific event regions.
    \item \textbf{$\mathcal{N}$ - Speaker counter:} 
    
    This attribute identifies the number of concurrent speakers at each frame and contains three event values: non-speech, single-speaker speech, and overlapped speech. 
    \item \textbf{$\mathcal{K}$ - Keynote speaker:} 
    
    This attribute focuses on identifying the keynote speaker. It contains one event value to represent the person who talks most. 

Identifying the keynote speaker is crucial for real-world applications. By leveraging both keynote and speaker counter prompts, user can differentiate between speech segments belonging to the keynote speaker and others.
\end{itemize}

\begin{figure}[!t]
    \centering
      \includegraphics[scale=0.63]{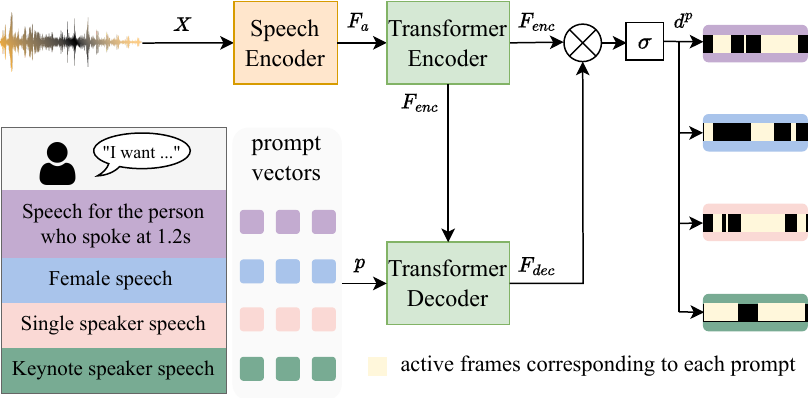}
    \caption{The overview of our prompt-driven target speech diarization framework. It takes audio and prompt vectors according to the user intention as inputs, and outputs target event regions aligned with the prompt vectors. $\otimes$ and $\sigma$ represent the dot product and sigmoid operation, respectively.}
    \label{fig:overview}
    \vspace{-2em}
\end{figure}

\vspace{-1.5em}
\subsection{Proposed framework: Prompt-driven Target Speech Diarization~(PTSD)}

To solve the task we formulated in the previous section, we proposed a framework called Prompt-driven Target Speech Diarization~(PTSD). In this framework, we modeled each semantic value information as a prompt vector $p \in \mathbb{R}^{1 \times D}$. Such a prompt-style framework~\cite{kirillov2023segment,lester2021power} can be the basis for constructing a versatile and flexible system. 

In this paper, we denoted the audio input as $X$ and formulated the target event region aligned with each semantic value as a binary sequence $y \in \{0, 1\}^{1 \times T}$, where $1$ represents the existence of target event and $0$ represents the absence. $T$ is the number of audio frame. As depicted in Figure~\ref{fig:overview}, PTSD compromises a speech encoder and a transformer encoder-decoder. The speech encoder first maps the input audio $X$ to a feature sequence $F_a \in \mathbb{R}^{T \times D}$. Then, the transformer encoder-decoder takes $F_a$ and prompt vector $p$ as input and outputs the prediction for target event related to $p$. Specifically, the transformer encoder takes $F_a$ as input and outputs the frame-level speech representation $F_{enc} \in \mathbb{R}^{T \times D}$. The transformer decoder takes prompt vector $p$ and $F_{enc}$ as inputs, and output $F_{dec} \in \mathbb{R}^{1 \times D}$. Finally, we performed a dot product operation between the decoder output $F_{dec}$ and the encoder output $F_{enc}$ and applied a sigmoid operation to get the prediction sequence $\mathrm{d}^p \in (0, 1)^{1 \times T}$. The value of $\mathrm{d}^p$ denotes the target event occurrence probability at each frame. Notably, our framework can support one or multiple prompt vectors at the same and output their associated target events regions accordingly.

%In this section, we introduce the architecture of our proposed Prompt-driven Target Speech Diarization~(PTSD).  As depicted in Figure~\ref{fig:overview}, \yidi{PTSD compromises a speech encoder and a transformer encoder-decoder. The inputs of PTSD include audio sequence $X$ and prompt vectors $p \in \mathbb{R}^{N \times D}$, which are the target event values that we have mentioned before. Here $D$ and $N$ is the feature dimension and the number of target event values, respectively.  The speech encoder processes the input audio to obtain the frame-level speech representations $F_a \in \mathbb{R}^{T \times D}$. $T$ is the number of audio frame. Then $F_a$ and prompt query vectors $p \in \mathbb{R}^{N \times D}$ are fed into the transformer encoder-decoder. 
%Then we perform a dot product operation between the decoder output $F_{dec} \in \mathbb{R}^{N \times D}$ and the encoder output $F_{enc} \in \mathbb{R}^{T \times D}$. By adding a sigmoid operation, the final results $\mathrm{d}^p \in \mathbb{R}^{N \times T}$ are the probability sequences to determine whether each frame matches the target event values:
\iffalse
\begin{equation}
\label{eq:main}
    \{\mathrm{d}_t^p\}^T_{t=1} = \left\{
	\begin{aligned}
		1 & , t^{th}~\text{frame matches the event value}\\
		0 & , t^{th}~\text{frame doesn't match the event value}
	\end{aligned}
	\right.
\end{equation}
\fi

\textbf{Speech encoder.} In our framework, we employed a pre-trained WavLM encoder~\cite{chen2022wavlm} as the speech encoder to obtain frame-level representations $F_a$. With consideration for the trade-off between computational efficiency and speech information, we utilized the first three layers WavLM encoder, freezing them during our training process. The WavLM encoder~\cite{chen2022wavlm} was designed to learn universal speech representations from vast amounts of unlabeled speech data, ensuring the universality and robustness of the frame-level audio representations.

\textbf{Prompt query vectors.} 
Each target event value is associated with its own prompt query vector. By switching between different prompts, our framework can accurately detect distinct event regions corresponding to each prompt query vector $p$.

For prompts belonging to the timestamped speaker attribute, we extracted a single vector from the temporal axis of the frame-level representation $F_a$ using the provided particular frame index for the timestamped speaker. This extracted vector serves as the prompt query vector of this timestamped speaker. For prompts related to gender attribute, we used two learnable embeddings to present male and female semantic values separately during the training stage. Similarly, for other attributes such as $\mathcal{N}$ and $\mathcal{K}$, we also employed learnable embeddings to provide information for each semantic value. 

\textbf{Transformer encoder-decoder.} 
We employed the transformer encoder-decoder architecture as introduced in~\cite{vaswani2017attention}.
Leveraging the power of self-attention and cross-attention mechanisms, the transformer encoder-decoder excels in capturing intricate temporal patterns in the audio data and aligning them with the relevant prompts. This synergy enables our model to precisely identify and diarize target event regions, making it a robust and adaptable solution for our task.

\textbf{Loss function.}
The learning targets of our framework are frame-wise binary ground truth labels $y \in \{0,1\}^{N \times T}$ of $N$ target events.
For each target event, we utilized binary cross-entropy loss to train our model, as defined in Equation~\ref{eq:loss}. ${d}_t$ and $y_t$ represent the predicted and ground truth labels of a specific target event for the $t^{th}$ audio frame, where $t \in [1, T]$. The loss function is designed to minimize the difference between predicted and ground truth labels, encouraging our model to accurately detect target event activities.
\vspace{-1.em}
\begin{equation}
    \mathcal{L} = -\frac{1}{T} \sum_{t=1}^{T} (y_t \cdot log d_t) + (1-y_t) \cdot log (1- d_t)
    \label{eq:loss}
\end{equation}
\vspace{-2.5em}
\section{Experiments}
In this section, we detailed the datasets, evaluation metrics, and experimental setup used to evaluate the proposed PTSD framework.
\vspace{-2.em}
\subsection{Dataset}
\label{sec:}
\vspace{-0.5em}
Given that real-world speech datasets cannot provide all the required groundtruth labels according to the semantic attributes introduced in Section \ref{ssec:task_formulation}, we followed the recipe\footnote{https://github.com/BUTSpeechFIT/EEND\_dataprep/} proposed in~\cite{landini2023multi} to simulate 2-, 3-, and 4-speaker datasets from Librispeech~\cite{panayotov2015librispeech}. To create datasets that closely resemble real-world conversations, we utilized conversation statistics from the DIHARD II development set~\cite{ryant2019second} to generate 1000 hours of audio for each sim2spk, sim3spk, and sim4spk dataset.

\vspace{-1.em}
\subsection{Evaluation metric}
\vspace{-0.5em}
For target speech diarization evaluation, we  primarily employed three metrics: accurate precision~(AP), area under the receiver operating characteristic~(AUC), and equal error rate~(EER) based on the implementation from sklearn package.

% We also report the diarization error rate (DER) which is the sum of miss speech~(MS), false alarm~(FA) and speaker confusion~(SC) when our framework works as diarization-related task.
\vspace{-1em}
\subsection{Implementation details}
\vspace{-0.5em}
The proposed PSTD framework was implemented using PyTorch and optimized with the Adam optimizer. We set the initial learning rate to $10^{-4}$ and decrease it by 5\% for each epoch. The dimension $D$ of audio feature $F_a$ and prompt query vector $p$ were both set to 256. For both transformer encoder and decoder structure, 4-layer transformer with 8 attention heads was applied. 
To ensure the robustness of our system, we conducted experiments using inputs of diverse lengths, spanning from 20 to 60 seconds during the training phase. For validation purposes, all inputs were segmented into 40-second chunks for simplicity.
\vspace{-1em}
\section{Results and Analysis}
\label{sec:res}
\vspace{-0.5em}
In this section, we presented a comprehensive demonstration of our framework's performance across heterogeneous prompts to show the feasibility of our proposed target speech diarization task.
Furthermore, our framework's applicability can be extended to target speaker activity detection, concurrent speaker counting and gender diarization tasks, each aligned with specific semantic attribute.
Moreover, we conducted a comparative analysis with the specialists model to evaluate the effectiveness of our approach.
\vspace{-1.em}
\subsection{Overall analysis of PTSD}
In this section, we conducted the training phase of our framework using the sim2spk, sim3spk, and sim4spk datasets. 
The performances are depicted in Table~\ref{tab1}, showcasing the AP, AUC, EER results for the prompts from four different semantic attributes. 
Notably, in the case of sim4spk, our model can support ten prompts vectors to specify ten target events regions simultaneously, comprising four timestamped speakers~($\mathcal{T}$), two under gender~($\mathcal{G}$) (related to female and male), three under speaker counter attribute~($\mathcal{N}$) (related to non-speech, single speaker speech, overlapped speech) and one under keynote speaker attribute~($\mathcal{K}$).

Impressively, all the AP and AUC values surpass 95\%, and EER values are under 7\%. 
These results demonstrate that our PTSD framework can accurately identify the desired event regions guided by provided prompt query vectors. 
In this way, we can switch the prompt query vector according to the user intention, which is flexible and powerful.
\vspace{-0.8em}
\subsection{PTSD with specific semantic attribute}
The results from the previous section indicate that we can utilize a single PTSD model by switching between different prompts to detect various speech events, and it has demonstrated commendable overall performance. To further evaluate PTSD, in this section, we compared PTSD with specialized models for different sub-tasks with specific semantic attribute.

\begin{table}[!t]
\centering
\caption{Overall analysis of PTSD system on the sim2spk, sim3spk and sim4spk datasets across diverse semantic attributes.}
\label{tab1}
%\begin{spacing}{1}
\begin{tabular}{c|cccc}
\hline
\textbf{Dataset}                  & \textbf{Attribute}             & \textbf{AP}~(\%)$\uparrow$ & \textbf{AUC}~(\%)$\uparrow$ & \textbf{EER}~(\%)$\downarrow$ \\ \hline
\multirow{4}{*}{\textbf{sim2spk}} & $\mathcal{T}$ &99.90  &99.91  &1.18         \\ 
                                  & $\mathcal{G}$ &95.65  &96.38  & 5.98         \\ 
                                  & $\mathcal{N}$ &99.84  &99.92  & 1.47        \\ 
                                  & $\mathcal{K}$ &99.91  &99.84  & 1.68               \\
%                                  & $\mathcal{E}$ &98.37  &98.58  & 5.71                 \\
                                  \hline
\multirow{4}{*}{\textbf{sim3spk}} & $\mathcal{T}$ &99.40  &99.65  & 2.80        \\ 
                                  & $\mathcal{G}$ &95.70  &96.46  &5.93         \\ 
                                  & $\mathcal{N}$ & 99.56 &99.77  & 2.41        \\ 
                                  & $\mathcal{K}$ &99.32  & 99.21 & 4.81                 \\
%                                  & $\mathcal{E}$ &  &  &                  \\
                                  \hline

\multirow{4}{*}{\textbf{sim4spk}} & $\mathcal{T}$ &98.88  &99.57  & 3.29        \\ 
                                  & $\mathcal{G}$ &96.70  &97.19  & 6.20        \\ 
                                  & $\mathcal{N}$ &99.53  &99.76  &2.51         \\ 
                                  & $\mathcal{K}$ &98.75  &98.88  &5.85                  \\
%                                  & $\mathcal{E}$ &  &  &                  \\
                                  \hline

\end{tabular}
%\end{spacing}
\vspace{-0.8em}
\end{table}

\vspace{-1em}
\subsubsection{PTSD~($\mathcal{T}$) v.s. TS-VAD}
Our framework also has the potential to perform the target speaker voice activity detection (TS-VAD) task when furnished with timestamp for each speaker, denoted as PTSD~($\mathcal{T}$). In the primitive TS-VAD paradigm~\cite{medennikov2020target}, an i-vector for the target speaker was provided to help the system to detect the speaking activity.

The key difference between TS-VAD and PTSD lies in the methods used to provide the speaker prior information: timestamped speaker prompt vectors in PTSD~($\mathcal{T}$) and speaker embeddings in TS-VAD.  

To ensure a fair comparison, we made modifications to the original TS-VAD model and named it as `mod. TS-VAD'. Specifically, we replaced the Bidirectional Long Short-Term Memory~(BLSTM) and MFCC used in the original TS-VAD with a four-layer transformer encoder and a three-layer WavLM encoder, respectively. Additionally, we employed the ECAPA-TDNN\footnote{https://github.com/TaoRuijie/ECAPA-TDNN} trained on a combination of VoxCeleb, CnCeleb~\cite{fan2020cn} and Alimeeting training sets, to extract accurate speaker embeddings for TS-VAD.

In PTSD, the timestamped speaker prompt vector corresponds to a temporal duration of 0.04 seconds (given that the length of the WavLM feature is 25 for one second) within the audio sequence. To ensure an objective comparison, we selected clean speech segments of length 1s, 2s, and 3s and employed them in the `mod. TS-VAD' system to extract speaker embedding. We provided these speaker priors for each input chunk to conduct the comparison experiments on Alimeeting and sim4spk datasets. Besides, since TS-VAD system was first introduced in a speaker diarization task~\cite{horiguchi2020end}, we used the diarization error rate (DER) as the evaluation metric for comparison.

The results, as shown in Table~\ref{tab2}, highlight that PTSD~($\mathcal{T}$) outperforms all three versions of TS-VAD on sim4spk dataset. Furthermore, PTSD also performs better than TS-VAD with 1s enrollment speech and achieves comparable results with 2s and 3s enrollment speech on Alimeeting dataset. This difference can be attributed to the ECAPA-TDNN encoder, which has been fine-tuned on the Alimeeting dataset. In contrast, our method does not rely on a specific speaker encoder, offering enhanced flexibility and convenience in this regard.
\vspace{-1.em}
\begin{table}[!htb]
\caption{The performance comparison between PTSD~($\mathcal{T}$) and TS-VAD on Alimeeting and sim4spk datasets.}
\label{tab2}
\vspace{-0.5em}
\begin{spacing}{1}
\begin{tabular}{p{2cm}<{\centering}|p{3.5cm}<{\centering}p{2cm}<{\centering}}
\hline
\textbf{Dataset}                  & \textbf{Method}             & \textbf{DER}~(\%)$\downarrow$  \\ \hline
\multirow{4}{*}{\textbf{Alimeeting}} & mod. TS-VAD~(with 1s ref) &12.63           \\ 
                                     & mod. TS-VAD~(with 2s ref) &10.22           \\ 
                                     & mod. TS-VAD~(with 3s ref) &\textbf{8.80}        \\ 
                                     & PTSD~($\mathcal{T}$) &11.40        \\ 
                                  \hline
\multirow{4}{*}{\textbf{sim4spk}} & mod. TS-VAD~(with 1s ref) & 12.63         \\ 
                                  & mod. TS-VAD~(with 2s ref) & 8.81         \\ 
                                  & mod. TS-VAD~(with 3s ref) & 7.03          \\ 
                                  & PTSD~($\mathcal{T}$) &\textbf{5.58}          \\ 
                                  \hline
\end{tabular}
\end{spacing}
\end{table}
\vspace{-0.5em}
\iffalse
\begin{table}[!h]
% \setlength{\tabcolsep}{6pt}
\caption{The performance comparison between Ours~($\mathcal{T}$) and other speaker diarization methods on CALLHOME 2-spk datasets.}
\label{tab3}
\begin{tabular}{p{3.cm}<{\centering}|p{0.9cm}<{\centering}p{0.9cm}<{\centering}p{0.8cm}<{\centering}p{0.8cm}<{\centering}}
\hline
\textbf{Method}                  & \textbf{DER}~(\%)             & \textbf{MS}~(\%) & \textbf{FA}~(\%) & \textbf{SC}~(\%) \\ \hline
x-vector clustering~\cite{horiguchi2020end} & 11.53 & & &           \\ 
BLSTM-EEND~\cite{fujita2019end} & 26.03 &&&           \\ 
SA-EEND~\cite{fujita2019end} & 9.54 &&&         \\ 
EEND-EDA~\cite{horiguchi2022encoder} &&&      \\
TS-VAD~\cite{medennikov2020target} & 9.51 &&&   \\
PTSD~($\mathcal{T}$) & 10.82  &3.09&7.27& 0.47     \\ \hline
\end{tabular}
\end{table}
\fi

It should be noting that the TS-VAD system cannot perform speaker diarization task independently. The authors in TS-VAD~\cite{medennikov2020target} leveraged a pre-trained diarization system to provide clean enrollment speech for each speaker, and then TS-VAD system was applied to detect the activity of each speaker. As an initial attempt, our experiments have demonstrated the feasibility of using timestamps as speaker enrollment priors to perform target speaker activity detection. Similarly, we can also extend our PTSD system in the same way as \cite{medennikov2020target} to complete the speaker diarization task. We will continue to explore the timestamp-based speaker diarization system in the future.

% However, similar to TS-VAD, which requires an extra clustering-based speaker diarization system to obtain clean enrollment speech for each speaker, our PTSD~($\mathcal{T}$) also requires additional information to assign timestamps for each speaker. As an initial attempt, our experiments has demonstrated the feasibility of using timestamps as speaker enrollment priors to effectively achieve traditional speaker diarization. We will continue to explore integrated traditional speaker diarization system with timestamps in future research.

\vspace{-1em}
\subsubsection{PTSD~($\mathcal{N}$) v.s. OSD}
PTSD can also be functioned as a three-class speaker counter, capable of estimating the number of concurrent speakers at each frame when we provide prompt vectors for non-speech, single speaker speech and overlapped speech simultaneously. We denote this mode as PTSD~($\mathcal{N}$).

We evaluated the performance of PTSD in the overlapped speech detection~(OSD) task on DIHARD II evaluation set. We used the overlapped speech prompt vector, which was initially trained on the sim4spk training set and further fine-tuned on the DIHARD II development set. Table~\ref{tab3} presents the comparison results between PTSD~($\mathcal{N}$) and previous two OSD models proposed by Jung et al~\cite{jung2021three} and Bullock et al~\cite{bullock2020overlap} respectively. PTSD~($\mathcal{N}$) achieves significantly better precision at $68.93\%$ and recall at $48.18\%$ compared to the specialized overlapped speech detection models on DIHARD II evaluation set.

\begin{table}[!h]
\setlength{\tabcolsep}{7pt}
\caption{Overlapped speech detection comparison results on DIHARD II evaluation set.}
\label{tab3}
\vspace{-0.5em}
\begin{tabular}{p{2.5cm}<{\centering}|p{2.2cm}<{\centering}p{2.2cm}<{\centering}}
\hline
\textbf{Method}  & \textbf{Precision}~(\%)$\uparrow$ & \textbf{Recall}~(\%)$\uparrow$ \\ \hline
Bullock et al.~\cite{bullock2020overlap} & 64.50 & 26.70  \\ 
Jung et al.~\cite{jung2021three} &  66.48 & 32.22     \\ 
PTSD~($\mathcal{N}$) & \textbf{68.93}  & \textbf{48.18}      \\ 
                                  \hline
% \multirow{2}{*}{\textbf{DH3 dev}} & Jung et al.~\cite{jung2021three} & 90.00  & 46.09       \\ 
%                                   & PTSD~($\mathcal{N}$) & 88.22 & 79.69        \\ 
%                                   \hline
\end{tabular}
\vspace{-1.5em}
\end{table}
\vspace{-1.em}

\subsubsection{Gender diarization: PTSD~($\mathcal{G}$)}
When we provide both female and male prompts, PTSD~($\mathcal{G}$) can perform for gender diarization for the first attempt to answer the question: ``which gender appeared when''.
We implemented the `baseline1' using WavLM speech encoder and ECAPA-TDNN encoder followed by fully connection layer as frame-wise binary classification. 
The second baseline we implemented is using WavLM speech encoder and transformer encoder with fully connection layer, denoted as `baseline2'. As shown in Table~\ref{tab4}, PTSD~($\mathcal{G}$) can get better performance than two baselines. At the same time, our framework can obtain more flexible prompt-driven outputs with transformer decoder structure.
\vspace{-0.5em}
\begin{table}[!h]
\setlength{\tabcolsep}{7pt}
\caption{Gender diarization on sim4spk datasets.}
\label{tab4}
\vspace{-0.5em}
\begin{tabular}{p{1.3cm}<{\centering}|p{1.2cm}<{\centering}p{1.2cm}<{\centering}p{1.2cm}<{\centering}p{1.2cm}<{\centering}}
\hline
\textbf{Method}   & \textbf{AP}~(\%)$\uparrow$ & \textbf{AUC}~(\%)$\uparrow$ & \textbf{EER}~(\%)$\downarrow$& \textbf{DER}~(\%)$\downarrow$ \\ \hline
baseline1 & 96.85 & 97.85 & 5.50 & 8.45     \\ 
baseline2 & 97.31 & 97.57 & 6.34 & 9.04      \\ 
    PTSD~($\mathcal{G}$) & \textbf{98.17} & \textbf{98.39} & \textbf{5.13} & \textbf{7.75}        \\ \hline
\end{tabular}
\vspace{-0.5em}
\end{table}
\vspace{-1.5em}
\subsection{Discussion and future work}
We believe that the attention mechanism plays a crucial role in making our framework operate effectively. Taking PTSD~($\mathcal{T}$) as an example, the attention structure combines information from nearby audio frames into the specific frame, thereby enriching the speaker-related information.

In the future, our research will progress toward integrating natural language commands and supporting various prompt forms. This expansion aims to improve the input query's effectiveness by adopting a multi-modal approach, which should enhance the system's adaptability and versatility in real-world applications.

\vspace{-0.5em}
\section{Conclusion}
\vspace{-0.5em}
In this paper, we have introduced an innovative task called "target speech diarization", aimed at distinguishing diverse speech events from various perspectives. This task mimics how humans naturally engage with audio in their daily lives. Additionally, we have proposed a framework, Prompt-driven Target Speech Diarization (PTSD), to replicate the multi-dimensional auditory comprehension process observed in humans. 
We have developed specific prompts for each target event, allowing us to switch between different functional modes. 
Our model's performance across various semantic attributes and the subsequent comparison with specialized models have demonstrated the superior performance and flexibility of our approach.
Our study provides new insights for diraization-related tasks. More practical application scenarios can be designed based on this direction.
\vspace{-1.0em}
\section{Acknowledgements}
\vspace{-0.5em}
This work is supported by Huawei Noah's Ark Lab, Human Robot Collaborative AI under its AME Programmatic Funding Scheme (Project No. A18A2b0046), the National Natural Science Foundation of China (Grant No. 62271432), Shenzhen Science and Technology Research Fund (Fundamental Research Key Project Grant No. JCYJ20220818103001002), the Internal Project Fund from Shenzhen Research Institute of Big Data under Grant No. T00120220002, and the Advanced Research and Technology Innovation Centre (ARTIC), the National University of Singapore under Grant (project number: A0005947-21-00, project reference: ECT-RP2).

\clearpage
\balance
\footnotesize
\bibliographystyle{IEEEbib}
\bibliography{refs}
\end{document}